\documentclass[useAMS,usenatbib]{mn2e}
\usepackage{float,graphics,epsfig,array,amsmath,amssymb}
\usepackage{times}

\title
[X-ray reverberation in 1H\,0707--495]
{X-ray reverberation in 1H\,0707--495 revisited}

\author
[L.~Miller et~al.]
{L.~Miller$^{1}$
\thanks{E-mail: L.Miller@physics.ox.ac.uk},
T.~J.~Turner$^{2,3}$,
J.~N.~Reeves$^{4}$
and V.~Braito$^{5}$\\
$^{1}$Dept. of Physics, Oxford University, 
Denys Wilkinson Building, Keble Road, Oxford OX1 3RH, U.K.\\
$^{2}$Dept. of Physics, University of Maryland Baltimore County, Baltimore, MD 21250, U.S.A.\\
$^{3}$Astrophysics Science Division, NASA/GSFC, Greenbelt, MD 20771, U.S.A.\\
$^{4}$Astrophysics Group, School of Physical and Geographical Sciences, 
Keele University, Keele, Staffordshire ST5 8EH, U.K.\\
$^{5}$Dept. of Physics and Astronomy, University of Leicester, Leicester LE1 7RH, U.K.
}

\begin{document}

\pagerange{\pageref{firstpage}--\pageref{lastpage}} \pubyear{2010}

\maketitle

\label{firstpage}

\begin{abstract}
The narrow-line Seyfert\,1 galaxy 1H\,0707--495 has previously been identified
as showing time lags between flux variations in the soft- (0.3-1\,keV) and medium-energy (1-4\,keV)
X-ray bands that oscillate
between positive and negative values as a function of the frequency of
the mode of variation.  Here we measure and analyse the lags also between a harder X-ray
band (4-7.5\,keV) and the soft and medium bands, using existing {\em XMM-Newton} data,
and demonstrate that the entire spectrum
of lags, considering both the full energy range, 0.3-7.5\,keV,
and the full frequency range, $10^{-5} \la \nu \la 10^{-2}$\,Hz,
are inconsistent with previous claims of arising as reverberation
associated with the inner accretion disk.  Instead we demonstrate that a 
simple reverberation model, in which scattering or reflection is present in all X-ray bands,
explains the full set of lags without requiring any {\em ad hoc} explanation
for the time lag sign changes.  The range of time delays required to explain the
observed lags extends up to about $1800$\,s in the hard band.  
The results are consistent
with reverberation caused by scattering of
X-rays passing through an absorbing medium whose opacity decreases with increasing
energy and that partially-covers the source.  A high covering factor 
of absorbing and scattering circumnuclear material is inferred.
\end{abstract}


\section{Introduction}

1H\,0707--495 (z = 0.0411) is a narrow-line Seyfert\,1 galaxy,
notable for the presence of an unusually deep drop in its X-ray spectrum just
above 7\,keV \citep{boller02a}. The spectral drop has been modeled as
an edge whose energy varies between observations \citep{gallo04a} and
the overall spectral form can be explained using partial-covering
models \citep{boller02a,tanaka04a,gallo04a}.  However, \citet{fabian04a}
suggested instead that the spectral
drop may be the blue wing of a relativistically-broadened Fe
K$\alpha$ emission line. Recently, 
\citet[][hereafter F09]{fabian09a} and
\citet[][hereafter Z10]{zoghbi10a}
have claimed that the soft X-ray spectrum of 1H\,0707--495 contains 
strong, blurred Fe\,L line emission below 1\,keV. 
F09 and Z10 explain the flux-dependent spectral variability
as due to varying contributions from a powerlaw continuum 
and an ionized reflector with high
metallicity. To fit the spectrum, the reflected emission is extremely
relativistically redshifted and blurred, with contributions down
to 1.23 gravitational radii ($GM/c^2$) from an illuminated disk with 
reflected emissivity
$\propto r^{-7}$ (where $r$ is the radial coordinate from the black
hole). Such a conclusion is surprising, given the extreme black hole spin, 
the very steep reflectivity profile required to obtain detectable
emission from so close to the black hole, and the implied very small size of
illuminating source.  
However, F09 argue that the
combination of this spectral modelling with an analysis of the source's
rapid flux variations does provide evidence for emission from a few
tens of light-seconds of the event horizon of a rapidly-spinning black hole,
and hence it is important to critically assess that model and test it
with further analysis.

In addition to the spectral modelling, 
F09 and Z10 searched for and detected time lags between 
flux variations in two X-ray
energy bands, 0.3-1\,keV and 1-4\,keV.  Lags were analysed in the Fourier domain,
so that they could be derived as a function of the frequency of the mode of
flux variation. Those analyses revealed the
first clear detection of ``negative lags'' in the X-ray time series of an
active galactic nucleus (AGN), where negative is defined as meaning that flux variations
in the softer
energy band lag variations in the harder energy band.   
Negative lags were found at high frequencies 
($\nu > 6 \times 10^{-4}$\,Hz) of amplitude up to about 40\,s. Larger positive
lags were found at lower frequencies.  Positive lags are well-known in AGN \citep[e.g.][]{mchardy04a},
but a previous detection of a negative lag was not strongly significant \citep{mchardy07a}
and the measurement of both positive and negative lags in an AGN had not been 
clearly established prior to the work of F09 and Z10.
F09/Z10 interpreted the high-frequency negative lags 
qualitatively as a reverberation signal originating within a 
gravitational radius of the black hole event horizon, and argued that this interpretation was consistent
with their spectral model, with its large component of relativistically-blurred
reflected emission in the soft band.
Z10 recognised that the larger-amplitude, low-frequency positive lags could not arise
from such inner-disk reflection, and therefore
suggested that these lags might have
an origin other than reverberation.
However, \citet{miller10a} analysed the positive lags in the time series of
{\em Suzaku} X-ray flux variations of NGC\,4051, and recognised that the energy-
and frequency-dependence of those lags may be well-explained by reverberation from
substantially more distant material, a few thousand light-seconds from the illuminating
source, leading to the question of whether such a reverberation model is relevant
also to 1H\,0707--495.

In their timing analysis of 1H\,0707--495, F09 and Z10
did not report any measurement of lags for energies above 4\,keV, thereby losing
the diagnostic potential of measuring a band that contains Fe\,K emission.  This
is particularly important given that the F09/Z10 spectral model predicts that an even
larger fraction of reflected light should be present in the Fe\,K spectral region.
In this paper we show measurements of the lags at higher energy and investigate 
both whether the Z10 model can explain the full set of observed lag behaviours and
whether a reverberation model such as that inferred for NGC\,4051 
is able to explain the frequency-dependent positive and negative lags seen
in 1H\,0707--495.

\section{Observations and data reduction}

\begin{table}
\caption{Summary of observations used: epoch,
observation ID, start date, duration rounded to the nearest ks,
and mean 0.3-7.5\,keV count rate.
\label{table:obs}
}
\begin{tabular}{ccccc}
epoch & ID & start date & duration /ks & count rate /s$^{-1}$\\
\hline
O2 & 0148010301 & 13-10-2002 & 68 & $3.841 \pm .007$\\
O3 & 0506200301 & 14-05-2007 & 36 & $2.028 \pm .007$\\
O3 & 0506200401 & 06-07-2007 & 17 & $4.227 \pm .016$\\
O3 & 0506200501 & 20-06-2007 & 34 & $6.116 \pm .013$\\
O4 & 0511580101 & 29-01-2008 & 93 & $3.504 \pm .006$\\
O4 & 0511580201 & 31-01-2008 & 68 & $5.163 \pm .009$\\
O4 & 0511580301 & 02-02-2008 & 68 & $4.509 \pm .008$\\
O4 & 0511580401 & 04-02-2008 & 68 & $3.656 \pm .007$\\
\hline
\end{tabular}
\end{table}

{\it XMM-Newton} has observed  1H\,0707--495 during four epochs, 
2000 Oct (O1), 2002 Oct (O2), 2007 Jan-Feb (O3) and 2008 Jan (O4), where
the names in parentheses indicate the 
naming convention used by Z10.   
As we might worry that different behaviour arises in low or high flux states, we only
consider observations whose flux is comparable to that of the long O4 observations.
Thus, in this paper we present the detailed analysis of 
data from 2002-2008, excluding the first of the 2007 snapshots 
(see Table\,\ref{table:obs} and Z10).
 
The European Photon Imaging Camera (EPIC) observations were made with
the medium filter and with the prime full window mode in 2002 and the
prime large window mode in 2007/8.  
Data were processed using {\sc SAS v8.0.0} and {\sc HEAsoft v6.8}.  
Our analysis used only the pn data,
which offered the best S/N ratio of all EPIC data and which were free of
pile-up effects.  The pn events were filtered using
standard criteria and instrument patterns 0--4 were selected.  
Source events were extracted from circular regions of
radius 35$''$ centred on the source, and background events
sampled from regions on the same chip about a factor 
$5$ larger in area than the source cell.  The 2007/8 observations
suffered from high background, with some background
flaring.  As noted by Z10, the high background is
problematic owing to prominent emission lines 
from the electronics circuit board.  The geometry of the circuit board
results in the cleanest area being the central region
of the pn chip array, 
comprising about the inner $40$\% of the chip at the nominal aim point
\citep{lumb02a}.  However, with the target positioned in the clean
area, it is difficult to find a sufficiently large
area for similarly clean background extraction on the same pn chip. In
order to be able to use the largest possible background extraction
area while minimizing contamination by the
background emission lines, we applied a tight background screening to
the data and removed 
periods when the full-band pn background count rate exceeded 0.25 count\,s$^{-1}$.
After screening, the data 
had a background level $< 1$\% of the source count
rate in the 0.3-1\,keV and 1-4\,keV bands and 10-15\%
of the source count rate in the 4-7.5\,keV band.  The total
effective pn exposures were 68\,ks (O2), 87\,ks(O3) and 297\,ks(O4).
1H\,0707--495 ranged in flux from $<1$\,count\,s$^{-1}$ in the lowest observed
states (not analyzed here) to $\sim 10$\,count\,s$^{-1}$ in the higher states
observed with {\it XMM-Newton}.

\section{Time lag measurement}
F09 and Z10 have previously considered the time lags only between two
bands: a ``soft'' band (0.3-1.0\,keV) and a ``medium'' band (1.0-4.0\,keV).
We add a further ``hard'' band covering 4.0-7.5\,keV, where the upper energy
bound has been chosen to include the Fe\,K region but 
avoid the region of greatest residual background
contamination.

To measure time lags between these bands we follow the method of \citet{miller10a},
which uses a maximum-likelihood formalism to measure the best-fitting power spectral
densities (PSD) in each band and the cross-power spectral densities and time
lags between bands.  
To measure a PSD alone,
the PSD is defined as the Fourier transform of the autocorrelation
function, $\mathcal{A}(\tau)$, as in other methods, 
but rather than trying to invert directly the observed
autocorrelation function, instead we maximise the likelihood, $\mathcal{L}$, 
of the observed $\mathcal{A}(\tau)$,
\begin{equation}
\mathcal{L} = \frac{1}{(2\pi)^{N/2}\left | C \right |^{1/2}}
\exp \left [-\frac{1}{2}\Delta^T C^{-1} \Delta \right ],
\label{eqn:likel}
\end{equation}
for a time series containing $N$ data samples,
where $C$ is a $N \times N$ model covariance matrix, whose elements are given by the model's
$\mathcal{A}(\tau)$, $\Delta$ is the time series vector of fractional
fluctuations with length $N$ and $\Delta^T$ is its transpose.
The likelihood function assumes the process that generates the time series 
to be gaussian (or, equivalently, assumes that the
Fourier transform of the intrinsic time series, before being observationally sampled,
has uncorrelated phases): 
in principle this assumption
could be varied and replaced by a lognormal distribution, however there is no strong evidence in
this time series for non-gaussian behaviour, and the assumption of gaussianity is common to
alternative methods of PSD estimation \citep[e.g.][]{uttley02a}.  
To find the best-fitting PSD we compute the expected covariance matrix
for an initial set of model parameter values and iterate using the method of \citet*{bond98a}.
The model covariance matrix includes the effects of shot noise \citep[see][]{miller10a}.  
The model parameters
are the amplitudes of the PSD in discrete ranges of Fourier frequency (and in the case of
estimation of the cross-spectrum for a pair of time series, the model parameters also include
the time lags, see below).  

The method has the principal advantages over other approaches that
it is immune to gaps in the time series data or to uneven time sampling
with no need for any interpolation (because the likelihood function is only defined for the
observed time series); it
corrects rigorously for the presence of shot noise (including the effect of uncertainty
in the shot noise); and it allows rapid estimation of the uncertainties in the
PSDs and time lags, including estimation of the covariance between frequency bins.  
Crucially, although PSD uncertainties are not gaussian, and neighbouring frequencies have
correlated errors, because the likelihood is calculated in the time domain the PSD uncertainties
and their covariance
are straightforward to calculate using either of two methods.  First, as part of the 
likelihood iteration, we calculate the model Fisher matrix (see \citealt{bond98a} for full
details) and when the model is a good description of the data, the square-root of the diagonal elements of the
inverse Fisher matrix yield an estimate of the uncertainties on the PSD parameters.
Strictly, these estimates are lower limits to the true uncertainties (the Cram\'{e}r-Rao bound)
and do not allow for the correlations between parameters which are encoded in the off-diagonal
elements of the inverse Fisher matrix.  To address these issues we may instead step through
a set of values for 
each parameter and measure the likelihood ratio, $\mathcal{L}/\mathcal{L}_{\rm max}$,
with respect to the best-fitting model, allowing other parameter values to float in the maximisation
(a procedure similar to marginalisation over those other parameters).  In this case,
we expect $\chi^2=-2\log(\mathcal{L}/\mathcal{L}_{\rm max})$ to be distributed as $\chi^2$ with
one degree of freedom.  Thus 68\,percent confidence intervals are found when $\chi^2$ increases
by unity.  The second method is more time-consuming, of course.  The error estimation methods have been 
tested against Monte-Carlo realisations for the \citet{miller10a} observation of NGC\,4051 by
Miller (in preparation) and found to be in excellent agreement with the distributions obtained
from those realisations.

To measure time delays between two time series, we follow \citet{miller10a} and extend the likelihood
maximisation to include the cross-spectrum of the pair of time series, by including both time series
in the vector $\Delta$ in equation\,\ref{eqn:likel}, 
so that the maximisation process involves handling matrices of rank $2N$.
The time delays are then obtained from the phases of the cross-spectrum \citep[e.g.][]{nowak99a}.

Although in principle the multiple observations described
in Table\,\ref{table:obs} may be analysed as a single time series, in practice this
is computationally time-consuming at high time sampling as it requires the inversion
of several $2N \times 2N$ matrices, a process which becomes prohibitively expensive on desktop
computers for $2N \ga 6000$.  We therefore follow an approach equivalent to that
of Z10 in which we ignore the temporal cross-correlation between time samples 
from different orbits, but combine the likelihoods deduced from each observation to obtain an
overall best-fit set of PSD and time lag values.

\begin{figure}
\begin{center}
\resizebox{0.4\textwidth}{!}{
\rotatebox{-90}{
\includegraphics{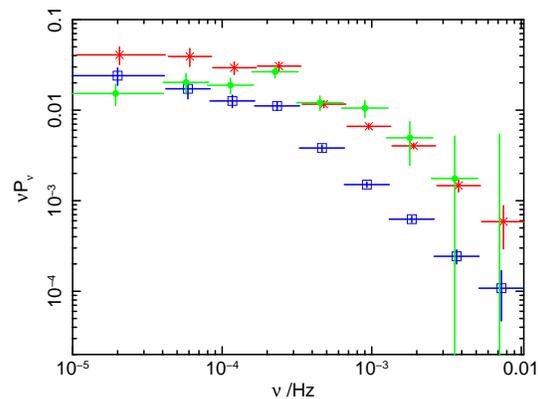}
}}
\caption{
Power spectral densities, $\nu P_{\nu}$, as a function of mode frequency $\nu$
for each of the three X-ray energy bands:
soft (open squares); medium (crosses) and hard (closed circles).
Vertical bars indicate 68\,percent confidence uncertainty, horizontal
bars show the range of frequencies included in each point.
Frequency values have been slightly offset between energy bands for clarity.
\label{fig:psd}
}
\end{center}
\end{figure}

\begin{figure*}
\resizebox{0.32\textwidth}{!}{
\rotatebox{0}{
\includegraphics{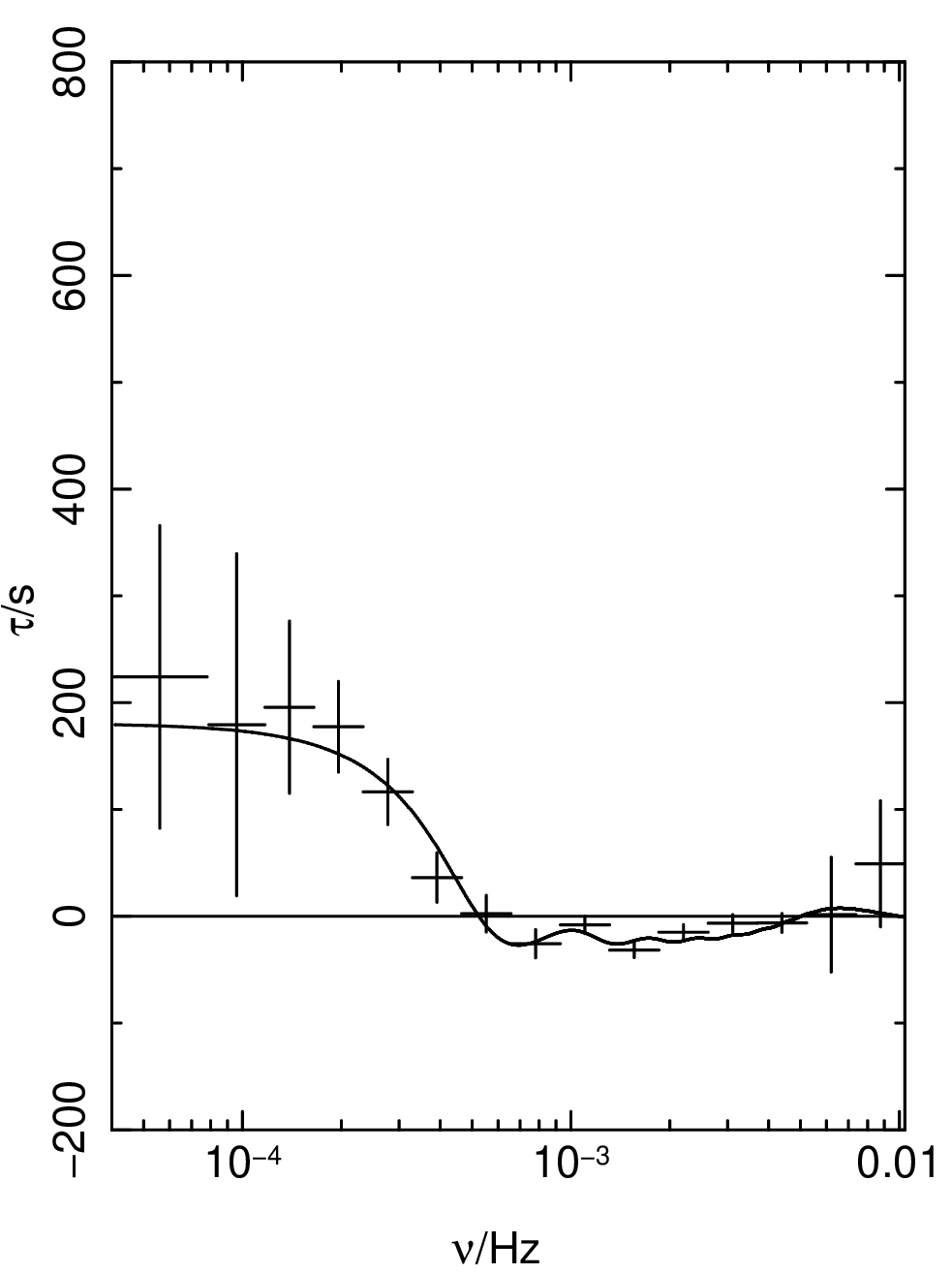}
}}
\resizebox{0.32\textwidth}{!}{
\rotatebox{0}{
\includegraphics{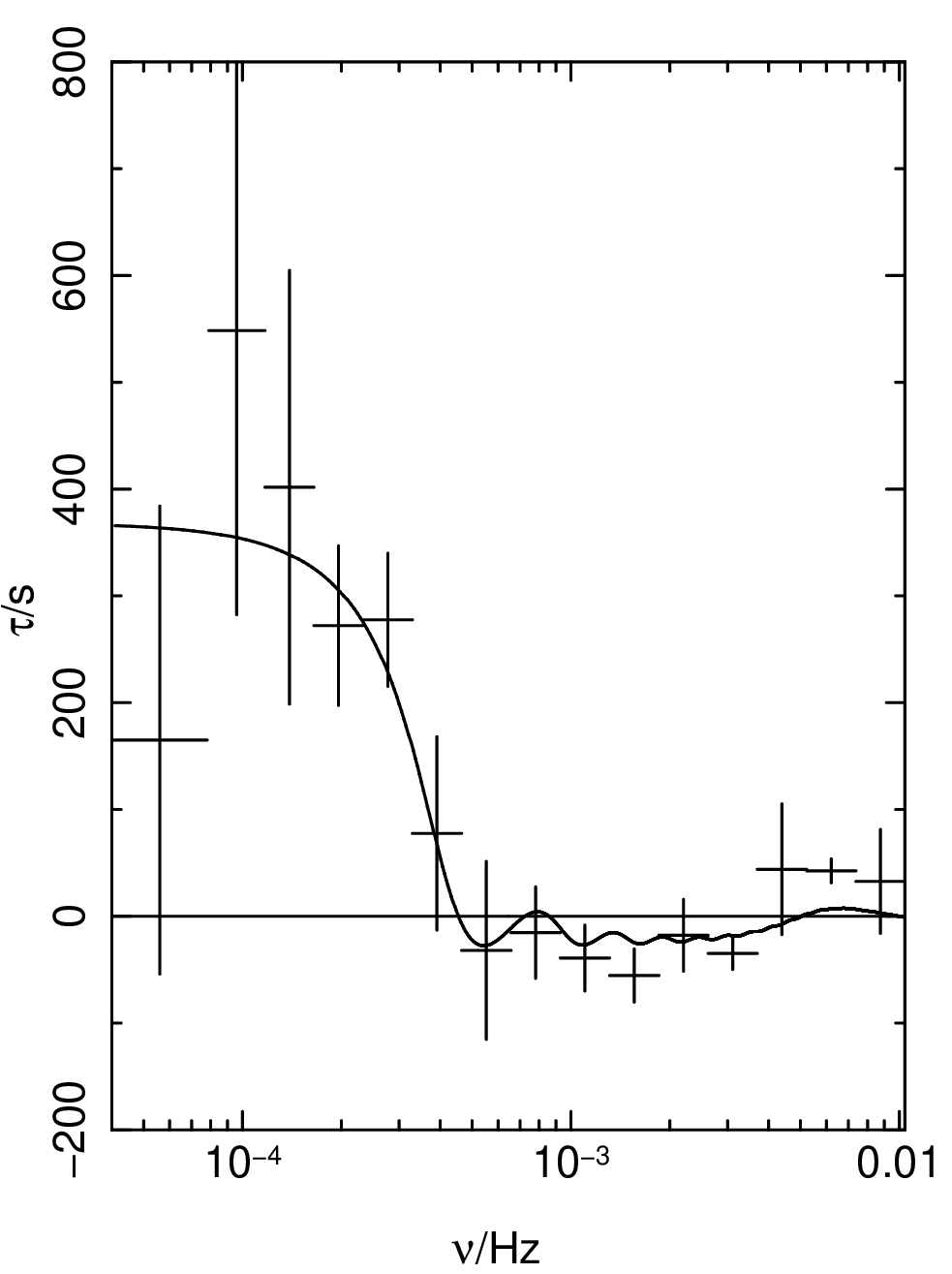}
}}
\resizebox{0.32\textwidth}{!}{
\rotatebox{0}{
\includegraphics{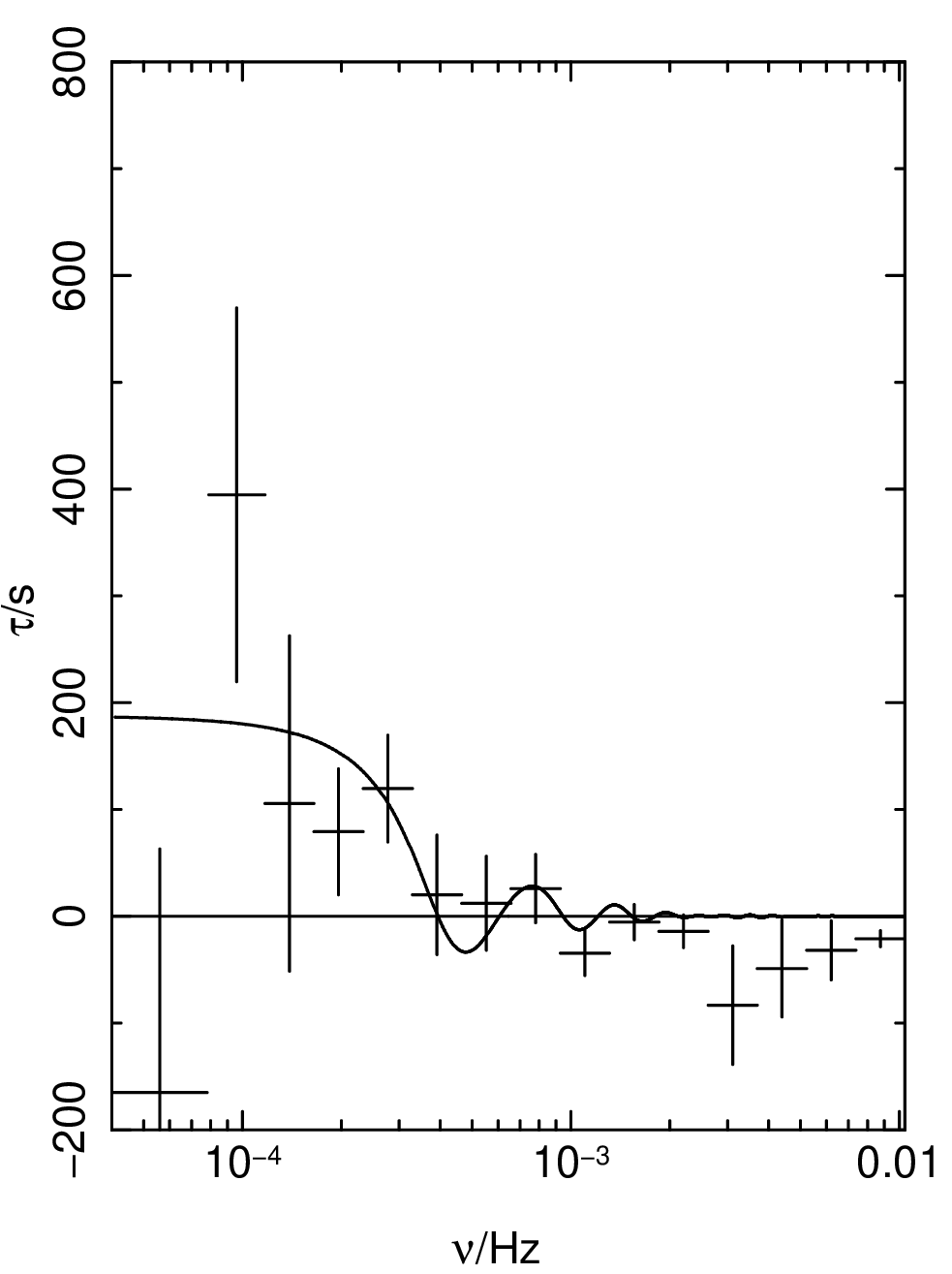}
}}
\caption{
Measured lag spectra (defined to be positive for the harder band lagging the softer band), 
shown as points with error bars, from the cross-band power
spectra of the medium v. soft bands (left panel), hard v. soft bands
(centre panel) and hard v. medium bands (right panel).  Vertical bars indicate 68\,percent
confidence regions from the stepped-likelihood method.
Horizontal bars indicate the range of frequencies used for each point.
Solid curves show the lag spectra expected from top-hat model transfer functions discussed in 
section\,\ref{sec:transferfunction}. 
\label{fig:lagspectrum}
}
\end{figure*}

Time series were generated with 48\,s sampling. PSDs
were measured for frequency bins of width $\Delta\log_{10}\nu=0.3$
for frequencies lower than the Nyquist frequency of 0.0104\,Hz.  The bin widths
were increased for the lowest frequency bins to ensure that their widths were at
least twice the lowest observable frequency of the longest individual dataset, so that
neighbouring frequency bins are largely uncorrelated.
Fig.\,\ref{fig:psd} shows the PSD estimates in each of the three energy bands.  
The vertical error bars show the expected 68\,percent confidence intervals expected from
the square-root of the diagonal elements of the inverse Fisher matrix, and hence do not
allow for covariance between parameters: we do not calculate stepped-likelihood uncertainties
as we make no further analysis of the PSDs in this paper.
The PSDs are
steep above $10^{-4}$\,Hz, although the harder bands have flatter PSDs
than the soft band (as found also in other AGN, e.g. NGC\,4051, \citealt{miller10a}).

Fig.\,\ref{fig:lagspectrum} shows the derived time lags as a function of frequency
(the ``lag spectrum'').  As we now need to sample rapid variations as a function of frequency,
a higher frequency resolution, $\Delta\log_{10}\nu=0.15$, was chosen, which also matches
more closely the resolution adopted by Z10.  Again, the frequency bin widths were increased
to maintain independence at low frequencies.
One further bin at lower frequency was included in the maximisation but is not shown in
Fig.\,\ref{fig:lagspectrum} as its uncertainty is too large to be useful.
The 68\,percent confidence region uncertainties shown were 
calculated by measuring the likelihood ratio obtained
when stepping through parameter values, as described above,
and thus account for the covariance with other parameters
and for the non-gaussianity introduced by transforming from the time domain to the Fourier-phase
domain.
As we have used the combined datasets O2, O3 (high-state) and O4, with rejection of
high-background segments,
and slightly different sampling from Z10, we do not expect the medium-soft lags to be 
identical to those of Z10.  There is however good general agreement, with maximum
positive lags of $237 \pm 72$\,s and maximum negative lags of $-32 \pm 5$\,s
at higher frequencies.  The uncertainties appear similar to those obtained by Z10.

We can see that
the hard-soft lags, not reported or discussed by F09 or Z10,
follow the same pattern with a larger amplitude of time lag.  The maximum positive
lag in this case is
$548 \pm 266$\,s, with the most statistically-significant individual positive lag being
$277 \pm 62$\,s.
The most negative lag has a value $-55 \pm 25$\,s.
The existence of several neighbouring lags with similar values further increases
the statistical significance of both the positive and negative lags.  The overall significance
of the negative lags is discussed further below.
The lags measured between medium and hard bands are not independent of the 
medium-soft and hard-soft
lags and are generally in reasonable agreement.  There is
some disagreement at the highest frequencies owing to
degeneracies in the fit, highlighting the difficulty of obtaining a unique measurement 
at high frequencies.

\section{Discussion}
\subsection{The interpretation of time lag spectra}
\label{sec:interpretation}
Before discussing in some detail the explanation of the observed lag spectra, there
are some general points to be made.
\begin{enumerate}
\item It has become common-place in X-ray timing analysis to inspect the time lag
spectrum and to consider each measured frequency as being in some sense independent
of the others.  This approach is flawed in the case where lags are caused by reverberation,
because in general we expect any reverberation signal that is restricted to some
finite range of time delays to produce lag spectral signals over the full measurable
range of frequencies.  The most extreme case would be for reverberation seen from a narrow
annulus on an accretion disk viewed face-on: in this case the transfer function is a delta-function
and its Fourier transform has amplitude and phase information distributed over all frequencies.
In the case of the analysis of 1H\,0707--495, any claim of a reverberation signal at high
frequencies should be accompanied by a demonstration of the expected time lag signatures
at all frequencies.
\item In reverberation models, 
the measured lag spectrum is a representation of the phases of the Fourier
transform of the transfer function.  Low-frequency modes have time lags in the same sense
as the reverberation lag (i.e., if the hard band contains more reflected emission than the soft
band, the low frequency modes have positive lags) but for modes whose time period is less than
the reverberation delay, the phase wraps around: phases are evaluated in the range
$-\pi < \phi \le \pi$ and so any apparent time lag, positive or negative,
is possible. A corollary statement is that 
high-frequency modes cannot generate large apparent time lags (i.e. any inferred time lag
at some frequency $\nu$ must satisfy $|\tau| < 1/2\nu$) regardless of the actual time delay.
\item Because in the X-ray band the signals are likely to be a combination of reflected,
time-delayed emission and directly-viewed emission, the measured time delay is always less
than the true time delay (i.e. the true time delay is diluted by the addition of the 
direct emission - see \citealt{miller10a}).  
\item In cross-correlating two bands and measuring the time delays between them,
there is likely to be reflected emission in both bands, and thus the inferred
transfer function is the cross-correlation of the individual transfer functions
of each band.
\item The existence of small time lags does not necessarily indicate a short light travel-time
between the source and the reflector, because reflection from material along the line of sight, at any distance 
from the source, has zero time delay with respect to directly-viewed emission, and slightly off-axis
material can result in arbitrarily small delays.  As an example, consider the transfer function
of a thin spherical shell of reflection, of radius $r$, isotropically surrounding a source.  The
transfer function is a uniform top-hat distribution with time lags $0 \le \tau \le 2r/c$
\citep{peterson93a}: the existence of lags close to zero do not imply that the reflecting material
has to be close to the source, and such a conclusion could not be reached unless a full model
reproducing the time lags for all modal frequencies were to be tested against the data.
\end{enumerate}

Bearing these points in mind, we shall first critically review the previous analyses
of the lag spectra of 1H\,0707--495, and then discuss a more viable model that explains all
the observed features.

\subsection{The inconsistency with inner-disk reflection}
The observed time lags are inconsistent with the hypothesis that they arise
from reflection from the inner accretion disk, as proposed by F09 and Z10,
for a number of reasons.  In this section we first discuss the hypothesis of F09/Z10,
that there are two independent lag-generating mechanisms that operate at different frequencies,
and we then move on to test consistency of the new hard-band lag measurements with the
F09/Z10 model.

\subsubsection{Constraints on two mechanisms of lag generation}
\label{sec:4.2.1}
Consider first the time lags discussed by F09 and Z10, between the soft 
and medium bands.
At low frequencies, $\nu \la 10^{-3.5}$\,Hz, the medium band lags the
soft band, which is inconsistent with the soft band containing delayed
reflection of the medium band.  The signature of
inner-disk reflection is supposed instead to be the small ($\sim 30$\,s) negative
lags  at high frequencies ($\nu \ga 10^{-3}$\,Hz).
This was claimed to place the reflecting region a few tens of light-seconds away from the
Fe\,L ionising source, traced by the medium band.  But if there are longer-timescale
fluctuations in that source, as observed, these must also be delayed
in the same sense: i.e. there should be a negative lag of tens of seconds
at low frequencies.  This is not observed, instead positive lags of $\sim 200$\,s 
are seen.  
F09 and Z10 suggested these 
arise as a separate process, such as the propagation of fluctuations
through an accretion disk from outer to inner radii,
with the inner disk emission supposedly having a harder powerlaw spectrum
than the outer disk, so that the softer band varies sooner than the harder band \citep{arevalo06b}.
This scenario could only produce the observed lag spectra if 
(i) inward radial perturbations leading to variable emission existed and
(ii) radial dependence of the spectrum existed and
(iii) the softer outer disk regions
did not vary on short timescales, whereas the harder inner disk regions did, and 
(iv) the outer disk emission did not result in any soft-band reflection, 
implying the outer disk spectrum to be strongly cut off just at the Fe\,L ionisation edge, and
(v) despite having emission from a wide range of radii, nonetheless the reflected emission
had a radial $r^{-7}$ dependence (F07, Z10).  Such a set of requirements is not necessary
if a more straightforward explanation is adopted (section\,\ref{sec:transferfunction}).
These considerations motivate us to carry out further evaluation of the viability
of the Z10 model.

\subsubsection{Testing the model for consistency with the hard-band lags}
\label{sec:4.2.2}
The hard band (4-7.5\,keV) analysed here also shows significant time lags
with respect to the soft band, again in the sense that the hard band lags the 
soft band by up to $\sim 550$\,s at low frequencies, with negative lags 
at high frequencies.  However, the hard band contains the Fe\,K region and, in the
Z10/F09 model, should also show delays with respect to its ionising continuum.  
According to the model, most reflection occurs in
the hard band, then the soft band, then the medium band with the least (but non-zero)
reflection (see Fig.\,8 of Z10). 
We find that the hard-soft lag
spectrum has an almost identical form to that of the medium-soft lag spectrum, but
with lags of larger amplitude, despite
the expectation from the Z10 model that the hard band supposedly has more reflection than the soft band
and thus should lag the soft and medium bands at $\nu \ga 10^{-3}$\,Hz (i.e. the hard-soft and hard-medium
lags should be positive where the medium-soft lags are negative).

We have assessed the statistical significance of the disagreement with the model 
as follows.  Starting with the medium-soft lag spectrum, we find the mean lag of the six
frequency bins that have negative lags, i.e. $0.00066 < \nu < 0.0053$\,Hz, by fitting a
a single time lag value to that frequency range and evaluating its uncertainty using the
maximum-likelihood method.  All other bandpower amplitudes and time lags were allowed to
vary during the likelihood maximisation.
The 68\,percent confidence interval for that one parameter
is given by $\Delta\chi^2 = -2\log(\mathcal{L}/\mathcal{L}_{\rm max}) = 1$, which yields
a mean lag $\tau_{\rm medium-soft} = -16.4 \pm 2.4$\,s.  The mean lags for the same frequency
range in the other comparisons between wavebands are $\tau_{\rm hard-soft} = -37.0 \pm 9.0$\,s and
$\tau_{\rm hard-medium} = -14.7 \pm 8.7$\,s (note that these three lag values are not
statistically independent but are consistent with each other).  Thus all three lag spectra
are significantly negative in the frequency range of interest, counter to the expectation
of the Z10 model. 

To compare with the Z10 spectral model, we estimate the expected relation between the lags in
the three bands as follows.
In a waveband where both direct and reflected light is present,
with reflection fraction $f$,
the inferred lag, $\tau^\prime$, is diluted with respect to the intrinsic time lag to the reverberating
region, $\tau$, such that $\tau^\prime \simeq f\tau/(1+f)$ for $\omega\tau \ll \pi$ when considering
Fourier modes with angular frequency $\omega$
\citep{miller10a}.  Thus the
inferred lag between two such bands, designated $i$ and $j$, 
with reverberation signals from the same region, is given by 
\begin{equation}
\Delta\tau_{ij}^\prime \simeq \left(\frac{f_i}{1+f_i}-\frac{f_j}{1+f_j}\right ) \tau.
\label{eqn:lagrelation}
\end{equation}
We obtained values for the reflection fractions $f_{\rm soft}$, $f_{\rm medium}$, $f_{\rm hard}$
in each of the soft, medium and hard bands respectively by refitting to the data the 
model described by Z10, with reflection spectrum given by relativistically-blurred
{\sc reflionx} tables \citep{ross05a}\footnote{
We note that the model is not a good fit to the data over the entire energy range, in particular
there are significant residuals of 20\,percent in the Fe\,K emission region, as seen in Fig.\,8
of Z10, in addition to significant residuals below 5\,keV that Z10 argue might be an effect of additional
variations in ionisation.
The extreme blurring in the model has erased features around Fe\,K that are present
in the data.
However, the Z10 model requires such extreme blurring, so we retain that model in our test.}.
As the time series analysis was carried out on variations in
count rate and not flux (as this optimises the signal-to-noise in the timing analysis) the fractions
$f$ have been obtained as ratios of counts in the blurred reflection component to the counts in
the power-law component in each band.
The reflection fractions in the summed observations were
$f_{\rm soft}=1.60$, $f_{\rm medium}=0.57$ and $f_{\rm hard}=2.03$ in the soft, medium and hard bands respectively.
The formal statistical uncertainties on the model fits are very small, however we expect there to be
some possible variation in allowable models, and to allow for this modelling uncertainty we also
fitted the Z10 model to the data from each of the four separate orbits that make up the main O4 observation. 
The sets of four reflection fractions were found to be
$f_{\rm soft} = \{1.63,1.13,1.57,2.48\}$,
$f_{\rm medium} = \{0.58,0.40,0.57,0.91\}$,
$f_{\rm hard} = \{2.04,1.45,2.05,3.58\}$
in each band for each of the observation IDs 0511580101, 0511580201, 0511580301, 0511580401
respectively.  
Inserting these reflection fractions $f$ into equation\,\ref{eqn:lagrelation}, 
we can predict the
expected relation between the observed lags in each band for any assumed value of distance
of the reflector from the primary source.

\begin{figure}
\resizebox{0.22\textwidth}{!}{
\rotatebox{-90}{
\includegraphics{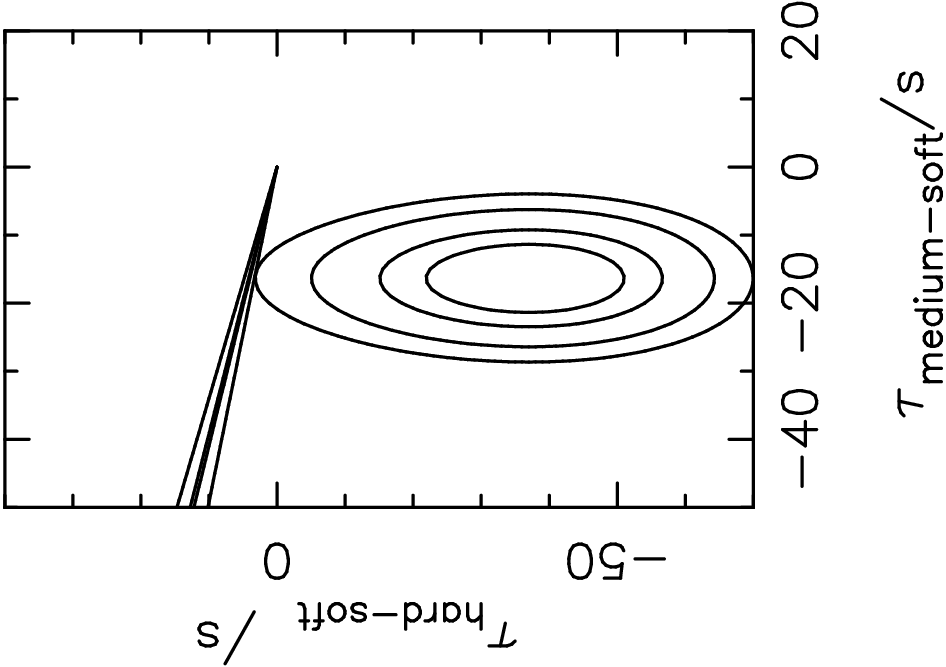}
}}
\hspace*{3mm}
\resizebox{0.22\textwidth}{!}{
\rotatebox{-90}{
\includegraphics{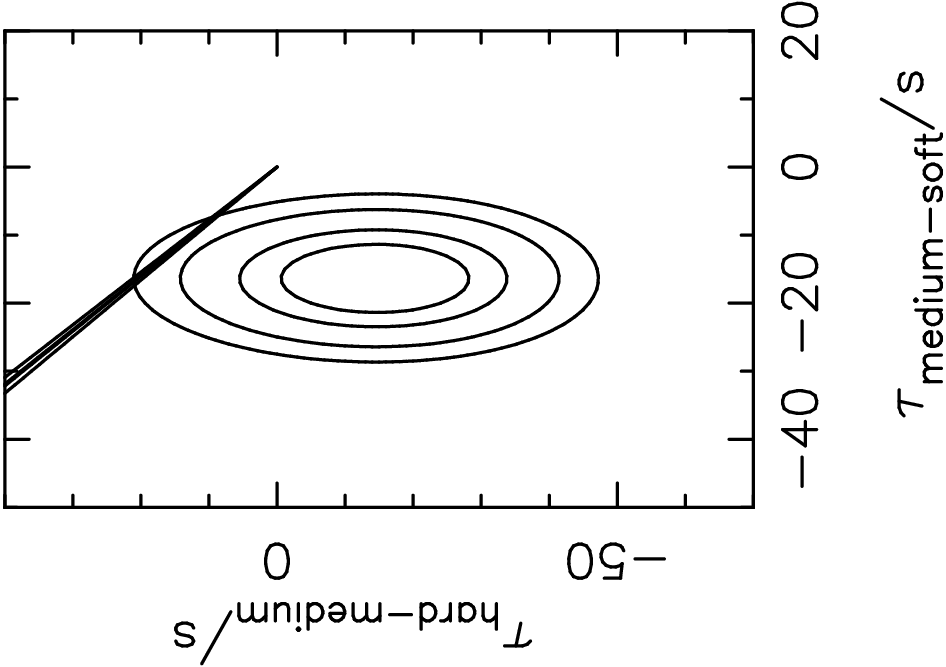}
}}
\caption{
Confidence regions for the mean lags in the frequency range 
$0.00066 < \nu < 0.0053$\,Hz between ({\em left})
hard-soft and medium-soft bands and ({\em right})
hard-medium and medium-soft bands.  Contours are shown 
corresponding to the 68\%, 90\%, 99\% and 99.9\% confidence
regions.  Straight lines indicate the relationship between
the lags in these bands expected from the Z10 model for each
of the four O4 observations
(only negative medium-soft lags are allowed in the Z10 model).
\label{fig:chisqplots}
}
\end{figure}

Fig.\,\ref{fig:chisqplots} shows these expected
relations and contours of $\Delta\chi^2$ obtained from the maximum-likelihood fitting,
comparing in turn the hard-soft lags and the hard-medium lags with the medium-soft lags.
Confidence region contours are shown at $\Delta\chi^2=2.3, 4.6, 9.2, 13.8$ which correspond
to the $68, 90, 99, 99.9$\,percent confidence regions for two parameters.
The failure of the hard-soft lags to have the opposite sign
to the medium-soft lags rules out
the Z10 model at $>99.9$\,percent confidence. The medium-soft lags
also rule out the model at $>99$\,percent confidence, although these two
measures are not independent.

\subsection{The reverberation transfer function} \label{sec:transferfunction}
We can use the considerations of section\,\ref{sec:interpretation} together with the 
new information from the hard band to infer what reverberation model could explain
the full set of available information.  Given that a reverberation explanation should also
seek to explain the large positive lags measured at low frequencies, we should identify 
the form of transfer function(s) that would produce the overall lag spectra.

It is important to note that the lag spectra derive from the phases of the Fourier
transform of the transfer function.  The amplitude information is carried in the 
cross-band power spectrum, but it is mixed in with the unknown underlying
cross-band spectrum, and so we have to rely on the phase information alone to inform us
about the transfer function.  Thus we cannot uniquely create a transfer function from the
lag information.  We can however consider some simple models that may have some link
to physical reality and that create the observed lag spectrum features.

\begin{figure}
\begin{center}
\resizebox{0.4\textwidth}{!}{
\rotatebox{-90}{
\includegraphics{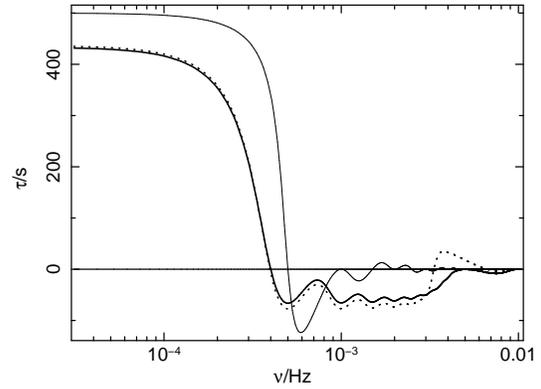}
}}
\caption{
Models illustrating oscillatory lag spectra: 
(a) {\em thin solid line} top-hat transfer function
in one band only;
(b) {\em thick solid line} top-hat transfer functions in two bands, both
transfer functions continuous from zero lag;
(c) {\em dotted line} as (b) but with no reflection delays shorter than 100\,s
(see text for details).
\label{fig:illustration}
}
\end{center}
\end{figure}

There are three effects that could lead to lag spectra of the form seen in 
Fig.\,\ref{fig:lagspectrum}.  First, transfer functions with sharp transitions
in general have oscillatory phases, and if the transfer function has a maximum away
from zero lag, the phase oscillations can go negative.  Such transfer
functions may be generated by clumpy reverberating material or material not isotropically
surrounding the source \citep{miller10a}.  As an example, consider a
transfer function that is a top hat in time delay, of width $\Delta t$ centred on delay $t_0$
and with intensity some fraction $f$ of the directly-viewed intensity. 
The lag spectrum,
$\tau(\omega)$ for angular frequency $\omega$, is given by
\begin{equation}
\tan\omega\tau = \frac{f\mathrm{sinc}(\omega\Delta t/2)\sin\omega t_0}{1+f\mathrm{sinc}(\omega\Delta t/2)\cos\omega t_0}
\end{equation}
An example is shown in Fig.\,\ref{fig:illustration}a for $\Delta t = t_0 = 1000$\,s, $f=1$.
Negative time lags are seen at high frequencies, although it is difficult to obtain 
negative lags over a wide frequency range from reverberation in one band only.
However, in section\,\ref{sec:interpretation}\,(iv) we noted that reverberation is likely to
be present at differing levels in both bands being cross-correlated.  
Fig.\,\ref{fig:illustration}b shows the lag spectrum expected for transfer functions
that are uniform from delay $t=0$ to $t_1$ where $t_1 = 1000$ and $200$\,s in each of two
bands, and with reflection fractions $f=1$ and $0.5$.  Now a broader range of negative
frequencies is seen.
Finally we consider the effect of the transfer functions having no reflection less than
some cutoff: Fig.\,\ref{fig:illustration}c shows the same transfer functions as (b)
but with no reflection at delays $t<100$\,s.  This allows additional positive oscillations
at high frequencies.

We can approximately 
fit this simple top-hat model to the measurements (Fig.\,\ref{fig:lagspectrum}).
The models shown were fitted jointly to the three lag spectra in two stages.
For speed, an initial fitting procedure minimised 
$\chi^2$ ignoring the covariance between parameters and the non-independence of the 
lag spectra.
Because the lag uncertainties are
non-gaussian, and because there is covariance between both lag values 
and cross-band power spectrum amplitudes, to estimate
the goodness-of-fit we then took this model and refitted the time series using the maximum
likelihood method, but with time lags fixed to the model values.  From this, the goodness-of-fit was 
evaluated as $\chi^2 = -2\log(\mathcal{L}/\mathcal{L}_{\rm max})$ where $\mathcal{L}$ was the
model likelihood and $\mathcal{L}_{\rm max}$ was the maximum likelihood found by allowing each
frequency bin to have an independent time lag (i.e. the values shown as discrete values in
Fig.\,\ref{fig:lagspectrum}).  
Considering just the hard-soft and medium-soft lags, to avoid the problem that all three lag spectra
are not statistically independent, 
the model shown had $\chi^2=23$ for 25 degrees of freedom,
a statistically-acceptable fit.  
The model fitted is one of the simplest possible models that we could have chosen, and the key
point of this exercise is that even this simple model generates well the large positive lags
at low frequencies, the sharp transition, and the extended range of high frequencies over which the lags
appear small but negative.  No such model has been demonstrated for the inner-disk reflection
hypothesis of F09/Z10, which was ruled out by the considerations of section\,\ref{sec:4.2.2}.

The best-fit top-hat start time was 100\,s, assumed common to all three bands.  
The maximum time delays for the hard, medium and soft
bands were 1800, 1450 and 150\,s respectively, and the ratios $f$ of flux in the reflected
component to flux in the directly-viewed component (which may itself be partially obscured)
were 0.67, 0.38 and 0.30 respectively.  We do not evaluate uncertainties on these parameter values
as this is computationally extremely expensive, 
requiring stepping through all parameter values and re-maximising
the likelihood in the time domain, and the model is almost certainly too simple to be taken too literally.
However, the values do indicate a possible physical origin for the different transfer functions
in the three bands.  The maximum time delay increases with energy: this is expected if X-rays are
scattered while travelling through a medium whose opacity decreases with increasing energy, because then
in the soft band X-ray photons cannot travel far through the medium before being absorbed, and any scattered
X-rays must have relatively short time delays, whereas hard X-ray photons can travel further and
create reverberation signals with longer delays.  In this case we would expect also to see a greater
fraction of scattered X-rays in the hard band, qualitatively in agreement with the inferred fractions.
However, the observed fractions are also affected by any absorption of the directly-seen radiation,
which cannot be determined from this analysis, and the predicted fraction of scattered light depends on
the unknown radial distribution of scattering and absorbing material,
making a more quantitative test difficult to achieve.
This inference implies that the reverberation transfer functions appear to arise from reflection or scattering  
from a distribution of material rather than from an unobscured reflector such as a naked accretion disk.
Given the high fractions of reflected light, we infer that either the reverberating material subtends a
large solid angle at the source, or that the directly-viewed emission is partially-obscured, or both.

According to this simple model, the reverberation in the soft band is characterised primarily by small delays,
$100 \la \tau \la 150$\,s.  However, as noted in section\,\ref{sec:interpretation}, the existence
of small lags cannot be taken as evidence for a short light travel time between source and reflector.
In the case of 1H\,0707--495, the model is only viable if the medium and hard bands have time delays up
to approximately 1800\,s.  Thus, even if one wished to assert that the soft-band reverberation were caused
by inner-disk reflection, one would require the medium and hard band reverberation to be produced at
significantly larger distances, with no physical explanation for that difference.  Such an
explanation would not allow models with these transfer functions
in which both soft-band Fe\,L and hard-band Fe\,K emission were reflected
from the same inner disk region.

\subsection{Spectral modelling and further work}
The aim of this letter has been to investigate the timing information for 1H\,0707--495 rather than to address
the spectral model.  The above analysis strongly indicates that there is a substantial component
of scattered light present in the spectrum.  Spectral modelling of this source to date has only considered either
extreme relativistically-blurred models with little obscuration (F09, Z10) or partially-covering
absorption models \citep{gallo04a} with no scattering.  Z10 claim to have ruled out the partial covering
models based on the weak observed strength of Fe\,K$\alpha$ emission, but we note that their argument
has already been discounted by \citet{miller09a} and \citet{yaqoob10a}.  
In fact, both sets of published models require
unattractively high values of Fe abundance, around 11 for the relativistically-blurred model (F09) and around
3 for the partial-covering model \citep{gallo04a}.  The indication that there is a substantial fraction
of X-rays scattered on passing through an absorbing medium means that models that include both
absorption and scattering are required.  The smooth spectral features in the soft band indicate that
the scattered emission is likely to be smoothed by Compton scattering as seen in the Monte-Carlo
radiative transfer models of transmission through a disk wind of \citet{sim08a, sim10a, sim10b}, and the presence
of a large amount of scattered light provides a natural way of obtaining a high equivalent
width in a broad, smooth emission feature, which is likely combined with strong absorption edges.
Further modelling such as that of \citeauthor{sim10a}\,(op. cit.) is required to make progress.

\section{Conclusions}
We have measured and analysed the time lags in 1H\,0707--495 between three X-ray bands:
the soft (0.3-1\,keV) and medium (1-4\,keV) bands discussed by F09 and Z10, and the hard band (4-7.5\,keV)
not previously considered by those authors.  Time lags have been analysed in the Fourier domain.
We can summarise our conclusions as follows.
\begin{enumerate}
\item The F09/Z10 model requires separate mechanisms for the positive lags at low frequencies
and the negative lags at high frequencies, and in Section\,\ref{sec:4.2.1} we have argued that
this imposes some special constraints on those mechanisms which may not be fulfilled in practice.
\item Consideration of the lags in all bands, soft, medium and hard, 
rules out models in which the inner accretion disk makes
a significant contribution at both Fe\,L and Fe\,K energies, and thus rules out the spectral model of F09/Z10.
In particular, the lags between the hard and medium bands and between the hard and
soft bands are also negative in the same range of frequencies that the lags between medium and
soft band are negative. Comparison between the observed lags and the lags predicted from the
spectral model rules out the Z10 model at confidence $>99.9$\,percent.
\item A simple model of reverberation, in which all X-ray bands have differing levels of reverberation,
provides a good quantitative description of the large positive lags at low frequencies and the high frequency
transition to negative lags, and fits the full lag spectra between all bands 
given the measurement uncertainties.
\item The harder X-ray bands appear to require longer maximum time lags than softer bands, consistent with
a geometry in which reverberation arises, not as reflection from a naked accretion disk, but from scattering
of radiation that is passing through partially-opaque material.  
Time delays at least as large as 1800\,s are required, placing a lower limit on the 
extent of the reverberating region of approximately 1000\,light-seconds, corresponding to 20--100 
gravitational radii for black hole mass in the range
$10^7$\,M$_\odot$ \citep{leighly04a} to $2 \times 10^6$\,M$_\odot$ (Z10, \citealt{zhou05a}) respectively.
\item Spectral models that include the effects of both absorption and Compton scattering are required.
\end{enumerate}
These results strengthen previous conclusions 
that X-ray observations of the inner regions of type-I AGN may be
dominated by the effects of circumnuclear material a few light-hours from the central source
\citep[e.g.][]{miller08a, risaliti09b, miller10a}. 
The material causes both scattering and absorption, and subtends a 
large solid angle at the source but with an anisotropic geometry.
Coupled with the evidence for energetic outflows \citep[e.g.][]{pounds09a, reeves09a}
it is likely that the reverberating material is associated with an accretion disk wind.

\vspace*{1ex}
\noindent
{\bf Acknowledgments}.
TJT acknowledges NASA grant NNX08AJ41G.
Observations were obtained with {\em XMM-Newton}, 
an ESA science mission with instruments and contributions directly 
funded by ESA Member States and NASA.
This research has made use of data obtained from the High Energy Astrophysics Science Archive 
Research Center (HEASARC), provided by NASA's Goddard Space Flight Center.

\bibliographystyle{mn2e}
\bibliography{xray_apr2010}

\label{lastpage}

\end{document}